\documentclass{iopjournal}

\usepackage{graphicx}
\usepackage{colordvi}
\usepackage{color,cite}
\usepackage{amsfonts}

\usepackage{amsmath,amssymb}

\usepackage{ragged2e}

\usepackage{caption}  

\begin{document}

\justifying

\title{Compound beams for direct experimental comparison of quantum operations}

\author{Kishore Thapliyal$^{1,2,*}$\orcid{0000-0002-4477-6041}, Jan Pe\v{r}ina Jr.$^{2,3}$, Pavel Pavl\'\i\v{c}ek$^3$,  and Anton\' \i n \v{C}ernoch$^{3}$}

\affil{$^1$Department of Physics,
University of Oslo, 0316 Oslo, Norway}

\affil{$^2$Joint Laboratory of
Optics, Faculty of Science, Palack\'{y} University, Czech Republic, 17. listopadu 12, 779~00 Olomouc, Czech Republic}

\affil{$^3$Joint Laboratory of Optics of Palack\'{y} University
and Institute of Physics of the Czech Academy of Sciences,
Institute of Physics of the Czech Academy of Sciences, 17.
listopadu 1154/50a, 779 00 Olomouc, Czech Republic}

\affil{$^*$Author to whom any correspondence should be addressed.}

\email{kishort@fys.uio.no}


\begin{abstract}
{\justifying Compound beams composed of simple experimental blocks
 (detected in simultaneous detection windows) that form
specific quantum-correlated structures are suggested  for
simulating the properties of different quantum operations used
for creating highly nonclassical and entangled  multi-mode
states needed in quantum communication, metrology, and information
protocols. Qualitative and quantitative comparison of multi-photon
addition and subtraction in compound multi-mode thermal as
well as sub-Poissonian beams and multi-mode twin beams with
their intensities extending over two orders in magnitude is
provided. Adding and subtracting up to twenty photocounts, optimal
conditions for the generation of experimental nonclassical states
are identified. In general, photon addition is identified as
advantageous over photon subtraction for the  multi-mode
thermal and sub-Poissonian beams: It induces (enhances) the
nonclassicality in the former (latter) state. Contrary to this,
photon subtraction outperforms photon addition in the
 multi-mode twin beams. Moreover, exploiting temporal
photon-pair correlations in compound twin beams when
post-selecting, nearly ideal experimental photon(s) addition is
demonstrated. The results apply for the beams with mean
photon number per mode much less than one. \par}
\end{abstract}

\section{Introduction}
The second quantum revolution harnesses nonclassical resources and
phenomena for technological advancement
\cite{dowling2003,deutsch2020}. Quantum, i.e. nonclassical, states
possessing non-positive Glauber-Sudarshan $P$-function
\cite{Glauber1963,Sudarshan1963} are essential to reach quantum
advantage. For instance, single-photon and entangled states are
ubiquitous nonclassical states for photonic quantum computation
\cite{kok2007}, sensing \cite{pirandola2018}, communication
\cite{Bouwmeester1997}, cryptography \cite{Gisin2002}, and imaging
\cite{defienne2024}. However, implementations of continuous
variable quantum information protocols often require non-Gaussian
states or operations \cite{Braunstein2005b}. Photon addition (PA)
\cite{Agarwal1991,Agarwal1992} and photon subtraction (PS)
\cite{Agarwal1992b} commonly applied to Gaussian states belong to
such non-Gaussian operations used, e.g., in entanglement
distillation \cite{Takahashi2010} and enhancement of the
nonclassicality.

They belong, together with photon catalysis
\cite{Lvovsky2002,Bartley2012} and other kinds of post-selection
schemes exploiting specific detection filters, to a broader class
of quantum operations applied in constructing highly nonclassical
states with specific properties. They are necessary for successful
implementations of numerous quantum-communication,
quantum-metrology, and in general quantum-information protocols.
Properties of the states generated this way have been addressed in
numerous papers using both theory and experiment. For instance, PA
and PS have recently been used to improve field's squeezing
\cite{Podoshvedov2025} and purity \cite{Zhang2025}. Similarly,
quantum error correction based on the cat codes
\cite{Cochrane1999} and GKP codes \cite{Gottesman2001,Konno2024}
require PS to generate superpositions of coherent states
\cite{Dakna1997,Ourjoumtsev2006,Gerrits2010}. Photonic simulations
of Maxwell's demon requires PS from uncorrelated and correlated
beams \cite{Vidrighin2016,Zanin2022}. PA followed by PS allows
noiseless amplification of coherent states \cite{Neset2025}.
Nevertheless, papers addressing direct comparison of effectiveness
and suitability of the generated properties for the states reached
by different operations are rare
\cite{Bartley2015,Barnett2018,Zhang2024}, especially from the
experimental point of view. The main reason is that different
light sources (or just their different parameters) are
experimentally convenient for different quantum operations. Here,
we suggest to use the so-called compound beams
\cite{PerinaJr2021b} as the original light sources ready to be
modified by quantum operations to overcome this obstacle. Compound
beams are built from different kinds of elementary building blocks
arising in the measurement on real optical fields with appropriate
properties. Compound beams adopt the properties of these real
beams and, on the top, they attain new properties originating in
the structure of the constructed beams.

At this point we would like to emphasize that the compound
beams are multi-mode contrary to the single-mode fields usually
used to study the properties of PS and PA. Nevertheless, provided
that the mean photon numbers per mode of these compound beams are
much less than one, the field properties originating in the
canonical commutation relations \cite{Vogel2001} can be omitted
and the compound beams provide reliable estimate of the properties
of multi-mode fields. We note that the technique of compound beams
is suggested to study the intensity properties of the fields. It
is not suitable for directly addressing the field phase properties
\cite{Perinova1998}, that are, however, not very interesting in
multi-mode fields.

As the compound beams are typically composed of greater number of
the elementary building blocks, the number of beams' modes is
greater. This means that their properties cannot be directly
compared to both theoretical and experimental results appropriate
for single-mode fields.
Nevertheless, various kinds of quantum fields retain their highly
nonclassical properties also in their multi-mode variants. Such
fields are then appealing for practical applications as their
generation in multi-mode form is usually much easier compared to
their single-mode form. The use of temporal multiplexed
measurements on weak twin beams (TWBs) in preparing compound beams
provides certain form of simulation of their statistical
properties as the realized states do not exist in real time.
However, there exists one-to-one mapping between the temporal and
spatial multiplexing that is realized when detecting multi-mode
fields by spatially resolved single-photon-sensitive detectors
like intensified CCD cameras or matrix single-photon-sensitive
detectors. In this case, the beams endowed with the properties of
compound beams exist in real experimental setups and they with
their useful properties are ready to serve for various tasks.
Specific kinds of such beams were analyzed in
\cite{PerinaJr2021,PerinaJr2021a,PerinaJr2024,PerinaJr2024a,Thapliyal2024,Thapliyal2024a}.
We note that the compound beams based on temporal multiplexing are
more versatile compared to their spatial counterparts as they
allow for the analysis of the beams in broader ranges of their
parameters including intensity, amount of the present noise, etc.
Here, the use of temporal multiplexing allows us to analyze
different kinds of the beams with their intensities differing by
two orders in magnitude.

Different kinds of highly-nonclassical beams constructed from weak
TWBs were experimentally investigated including those allowing for
the experimental investigations of bi- and tri-partite
entanglement in multi-mode Gaussian states
\cite{Adesso2007,Barasinski2023,Sudak2024,PerinaJr2025},
sub-Poissonianity of the beams generated by post-selection
\cite{Davidovich1996,Laurat2003,Iskhakov2016,PerinaJr2021b,PerinaJr2024a},
states with checkered-pattern photon-number distributions
\cite{PerinaJr2021a}, etc. To demonstrate the power of the
approach and usefulness and versatility of compound beams, we
apply them to directly compare the properties of the states
emerging in the quantum operations of PA and PS applied to three
important and frequently used kinds of optical fields: compound
 {multi-mode} twin beams,  {multi-mode} thermal states
(TSs), and sub-Poissonian states (SPSs) (for the used acronyms of
the analyzed fields, see Tab.~\ref{tab1}).
\begin{table}   
\centering {\setlength{\tabcolsep}{1.5pt}
\begin{tabular}{|c|c|c|c|}
\hline
Acronym  & Field  & Acronym  & Field \\
\hline \hline
TS & thermal state & PATS & photon-added thermal
state \\
 & & PSTS & photon-subtracted thermal state \\
\hline
SPS & sub-Poissonian state & PASPS & photon-added
sub-Poissonian state \\
 & & PSSPS & photon-subtracted sub-Poissonian state \\
\hline TWB &  twin beam & PATWB &  photon-added twin beam \\
 & & PSTWB &  photon-subtracted twin beam \\
\hline
\end{tabular}}
\caption{ {Acronyms of the fields analyzed in the text.}}
\label{tab1}
\end{table}

In general, experimental generation of nonclassical states
requires nonlinear optical processes, among which spontaneous
parametric down-conversion is the most prominent due to its
relatively high efficiency \cite{Mandel1995}. It provides an
entangled state endowed with quantum correlations originating in
photon pairing with common thermal photon-number statistics. Twin
beams composed of numerous signal-idler photon pairs render
sub-Poissonian state [TS] in the marginal signal beam
with [without] post-selecting photon-number-revolving measurement
in the idler beam \cite{Laurat2003}. TSs have photon number
fluctuations broader than its average photon number, in contrast
to SPSs. This means that TSs, SPSs, and TWBs can be obtained from
a common field originating in parametric down-conversion. From
this point of view, all three kinds of the fields share certain
common properties, that can directly be compared. The same holds
when PA and PS are considered. Experimental PA was performed
separately on TSs, SPSs \cite{Thapliyal2024a}, and also TWBs
\cite{PerinaJr2024a,Thapliyal2025c}. Similarly, PS was
successfully reported in two-mode squeezed vacuum state
\cite{MaganaLoaiza2019} and  {Schr\" odinger kitten states}
\cite{Ourjoumtsev2006}, multi-mode graph states
\cite{Ra2020,Walschaers2020}, multi-mode TSs and SPSs
\cite{Thapliyal2024a}, and multi-mode TWBs
\cite{Thapliyal2024,PerinaJr2024,Thapliyal2025c}. In these works
superiority of the states reached by PA and PS over the other
states was demonstrated in numerous applications, including
quantum metrology, imaging \cite{defienne2024}, and
entangled-photon virtual-state spectroscopy
\cite{Fei1997,Saleh1998,Svozilik2018,Svozilik2018a,Thapliyal2024,PerinaJr2024a}.

Thus, the processes of PA and PS have been extensively studied in
the literature, both theoretically and experimentally, across
various types of states and operating conditions, demonstrating a
range of advantages and practical applications. They strongly
depend on whether theoretical (ideal) advantages are analyzed or
the beneficial properties are investigated under specific
experimental conditions. To put these investigations on the same
footing and provide this way fair mutual comparison, we utilize
the above TWBs, TSs, and SPSs arising in the same source.
This fair comparison in a common experimental setup, involving
optical components with parameters typical for current stage of
technology development, is critical for competent choice of the
type of the state optimal for the use in a specific 
application.

\begin{figure}[t]   
 \centerline{(a) \includegraphics[width=0.7\hsize]{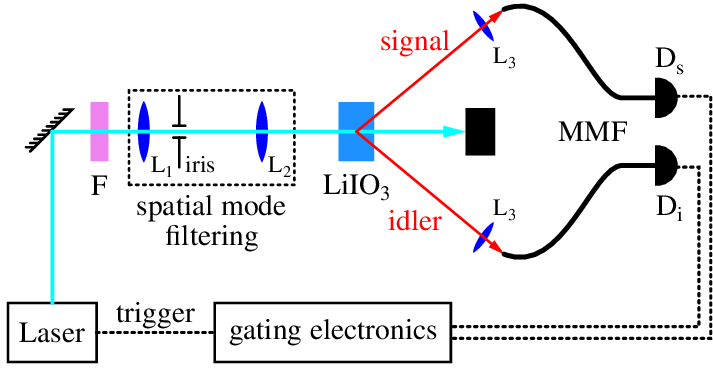}}
 \vspace{2mm}
 \centerline{\includegraphics[width=.95\hsize]{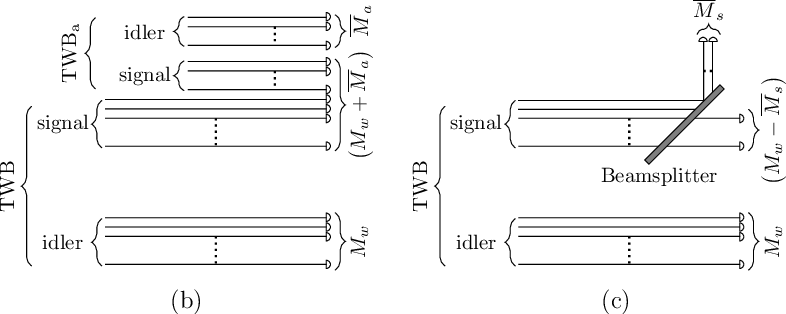}} 

 \caption{(a) Experimental setup involving nonlinear crystal LiIO$ {}_3 $
  as the source of photon pairs with the signal and idler photons
  propagating along different directions and synchronously detected by two APDs $\rm D_s$ and $\rm D_i$;
   {F - frequency filter; L$ {}_1 $, L$ {}_2 $, L$ {}_3 $ - lenses, MMF - multi-mode
  fiber.}
  Schematic diagrams for PA and PS in a TWB are shown in (b) and (c), respectively,  {using the synchronized signal and idler
  channels.}
  Whereas real PA requires an auxiliary TWB$_{\rm a}$ and
  post-selection, PS is realized on the signal beam by inserting
  a beam splitter with transmissivity $ T $ effectively realized by specific grouping of detection windows
  in the signal channel:  {$ M_{\rm w} $, $ \bar{M}_{\rm a} $, and $ \bar{M}_{\rm s} $ give in turn
  the number of detection windows in TWB, TWB$_{\rm a}$, and used for simulating the reflected beam at
  the beam splitter.}} 
\label{fig1}
\end{figure}

\section{Results}

\emph{Experimental setup ---} In our experimental setup, we
generated the compound TWBs with varying intensity by
concatenating different and increasing numbers of identical weak
photon-pair fields--- weak TWBs. These weak TWBs originated in the
experimental setup shown in Fig.~\ref{fig1}(a) in which a
nonlinear (LiIO$ {}_3 $) crystal emitted weak pulsed TWBs that were
detected by two single-photon sensitive avalanche photodiodes
(APDs) \cite{Hong1985,Hong1987,PerinaJr2021b}.
Type-I parametric down-conversion in a LiIO$_3$ nonlinear crystal
pumped by the third harmonic of an Nd-YAG laser at 355~nm and
repetition rate 2.5~kHz was used to generate weak TWBs at
degenerate photon wavelengths centered at 710~nm. The third
harmonic was spectrally cleaned by dichroic mirrors and 10-nm-wide
spectral filter. Spatial filtering of the pump beam by a $4f$
system composed of two lenses (L$_1$, L$_2$) with the focal
lengths of 20~mm and 50~mm and the 50-$ \mu $m-wide round aperture
was performed to improve quality of the beam. The APDs (Count-NIR
from Laser Components) with nominal detection efficiencies about
80~\% and very low dark count rates ($<50$~Hz) were used to detect
the generated beams. The signal and idler photons were spatially
filtered by coupling into multi-mode fibers using lenses L$_3$
(15.3~mm focal length, 5~mm clear aperture). Electronic signals
from the detectors were recorded simultaneously with the trigger
from the laser using the counting logic electronics and then
recorded. Pump pulses were 6.5~ns long and the signal and idler
photons were emitted together in 6.5-ns-long time windows. Time
delay between subsequent pulses was 4~ms. The laser power was
stabilized in a feed-back loop to keep stable operation of the
experiment over tens of hours. The average single-photon detection
rate was monitored during the measurement and the laser pump power
was adjusted whenever the detection rate changed by more than
5~\%.
Data obtained in consecutive detection windows were stored as two
synchronized channels, one providing the signal photocounts $
\tilde{c}_{\rm s}$, the other the idler photocounts $
\tilde{c}_{\rm i}$. The TWBs were sufficiently weak so that
either 0 or just 1 photocount arised in one detection window.
Concatenation of up to 200 detection windows offered us TWBs
containing up to approx.~80 mean photon pairs ready for
the PA or PS application. More details of the experimental setup
as well as the technique of compound TWBs can be found
in~\cite{PerinaJr2021b}. The method is versatile and large
numbers of available simple measurements on weak TWBs ($M_{\rm
m}=645 \times 10^6 $) allow to construct large measurement
ensembles for individual compound beams. These beams are analyzed
below as they change after addition and subtraction of up to 20
photocounts. Theoretical considerations and modeling of these
processes are provided in \cite{Thapliyal2024a}.

\emph{Processing of experimental data ---} Careful processing of
the detection data in two synchronized channels, schematically
illustrated in Fig.~\ref{fig1} and described in Supplemental
Material (SM) \cite{SM}, leaves us the measurement histograms $
f(c_{\rm s},c_{\rm i}; M_{\rm w}) $ corresponding to the detection
of a TWB with $ c_{\rm s} $ signal and $ c_{\rm i} $ idler
photocounts in neighbor $M_{\rm w}$ detection (double) windows.
In these histograms, detailed structure of a compound beam,
that reflects the way how the beam was constructed from the
elementary building blocks, is concealed. The original modes
present in the elementary building blocks disappear and their
photons now belong to the whole field. The field is then
characterized by the quantities derived from these histograms,
including their marginal and conditional histograms.
We note
that, in the experiment, the signal and the idler detection
channels have different properties and so, to suppress this
difference, we mutually exchange every second entry in the signal
and the idler channel. The marginal histogram $ f_{\rm th} $ then
characterizes a TS, i.e. $ f_{\rm th}(c_{\rm s}; M_{\rm
w})=\sum_{c_{\rm i}=0}^{\infty} f(c_{\rm s},c_{\rm i}; M_{\rm w})
$. Similarly, the histogram $ f_{\rm sP} $ belonging to an SPS
obtained from the TWB by post-selecting $c_{\rm i}$ idler
photocounts is obtained as $ f_{\rm sP}(c_{\rm s};c_{\rm i},
M_{\rm w}) = f(c_{\rm s},c_{\rm i}; M_{\rm w}) / \sum_{c_{\rm
s}=0}^{\infty} f(c_{\rm s},c_{\rm i}; M_{\rm w}) $. Below, we
consider SPSs for specific post-selecting $ c_{\rm i} $ that are
the most probable.

\emph{Groups of analyzed states ---} The application of PA and PS
on three prepared states TWBs, TSs, and SPSs leaves us with 9
groups of potentially nonclassical states that we systematically
denote as TWB, PATWB, PSTWB, TS, PATS, PSTS, SPS, PASPS, and
PSSPS, all of them available under the same experimental
conditions (acronyms are summarized in Tab.~\ref{tab1}).

 To experimentally perform PA, we take additional
$\overline{M}_{\rm a}$ detection double windows from the channels
to build an auxiliary TWB$_{\rm a}$, as illustrated in
Fig.~\ref{fig1}(b). The signal beam conditioned on the detection
of $\bar{c}_{\rm a}$ idler photocounts is then considered as a
realistic realization of the field with a specific number of
signal photons to be added. Adding this signal conditioned beam to
a TWB detected in $ M_{\rm w} $ double windows we arrive at a
PATWB with the histogram $f^{\rm a}(c_{\rm s},c_{\rm i};
\bar{c}_{\rm a},M_{\rm w},\overline{M}_{\rm a})$, whose signal
(idler) beam extends over $ M_{\rm w} +\overline{M}_{\rm a} $ ($
M_{\rm w} $) detection windows. Similarly we reveal PATSs and
PASPSs observed in $M_{\rm w}+\overline{M}_a$ detection windows
having the histograms $ f_{\rm th}^{\rm a}(c_{\rm s}; \bar{c}_{\rm
a},M_{\rm w},\overline{M}_{\rm a}) $ and $ f_{\rm sP}^{\rm
a}(c_{\rm s};c_{\rm i}, \bar{c}_{\rm a}, M_{\rm
w},\overline{M}_{\rm a}) $, respectively. On the other hand, PS is
implemented as a beam splitter that effectively splits the $
M_{\rm w} $ signal detection windows into $ M_{\rm w}
-\overline{M}_{\rm s} $ windows and $\overline{M}_{\rm s} $
windows that mimics the behavior of a beam splitter with
transmissivity $ T_{\rm s} = 1 -\overline{M}_s/M_{\rm w} $ [for
the scheme, see Fig.~\ref{fig1}(c)].  {We note that this
simulation of transfer of a compound beam through a beam splitter
has its direct real experimental realization when using real
multi-mode fields detected by an iCCD camera \cite{PerinaJr2021}.
In the experiment reported in~\cite{PerinaJr2021}, actual division
of a field input at a simulated beam splitter in fact occurs in a
nonlinear crystal that emits photon pairs into different
spatio-spectral modes.} Conditioning on detecting $\bar{c}_{\rm
s}$ photocounts in the auxiliary (reflected) signal beam, we
recover the histogram $f^{\rm s}(c_{\rm s},c_{\rm i}; \bar{c}_{\rm
s},M_{\rm w},\overline{M}_{\rm s})$ of the corresponding PSTWB, as
well as the histograms $ f_{\rm th}^{\rm s}(c_{\rm s};
\bar{c}_{\rm s},M_{\rm w},\overline{M}_{\rm s}) $ and $ f_{\rm
sP}^{\rm s}(c_{\rm s};c_{\rm i}, \bar{c}_{\rm s}, M_{\rm
w},\overline{M}_{\rm s}) $ of the marginal PSTS and post-selected
PSSPS, respectively. In both processes, for practical reasons, the
free numbers $\overline{M}_{\rm a}$ and $\overline{M}_{\rm s}$ of
detection windows are fixed such that the probability of PA or PS
success is maximal (see Tab.~\ref{tab2}  {and SM for details}).
\begin{table}   
 \caption{{\bf Probabilities of generating $
 \bar{c}_{\rm a} $-PA and $
 \bar{c}_{\rm s} $-PS states}${}^{\rm a}$}
\centering {\setlength{\tabcolsep}{1.5pt}
\begin{tabular}{|l | c | c | c |}
\hline
State    & $ \bar{c}_j =1$ & $ \bar{c}_j =5$ & $ \bar{c}_j =20$   \\
\hline
Photon addition ($j={\rm a}$) & 38.99 (38.03) & 18.60 (18.15) & 9.40 (9.16) \\
Photon subtraction ($j={\rm s}$) & 38.98 (38.50) & 18.59 (18.30) & 9.40  (9.20)  \\
\hline
\end{tabular}}
\\
${}^{\rm a}$  {Theoretical predictions given in \% are written
in parentheses that follow their experimental counterparts
(relative error $ 1\%$).}
\label{tab2}
\end{table}

\begin{figure}[t]  
 \centerline{\includegraphics[width=0.3\hsize]{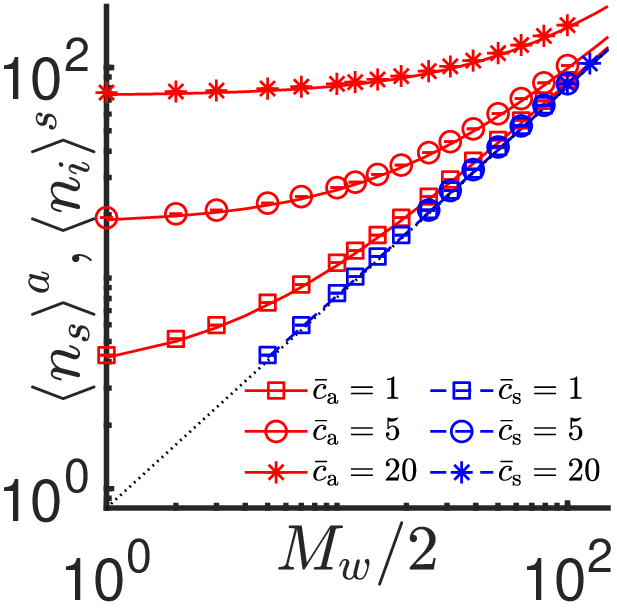}
 \hspace{3mm}
\includegraphics[width=0.3\hsize]{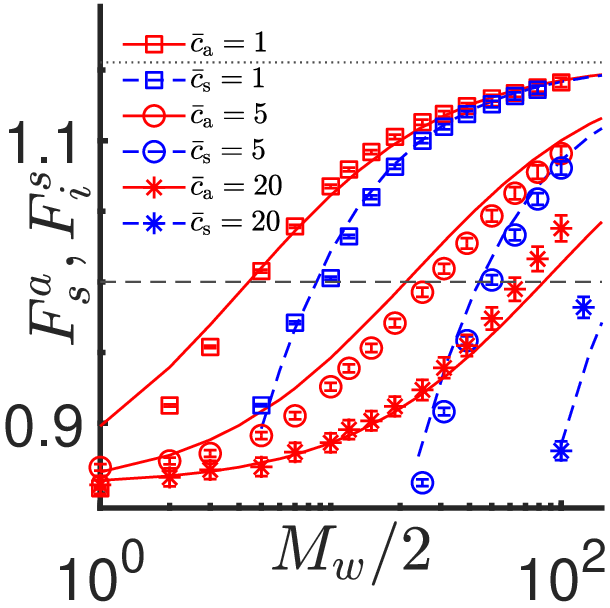}
 \hspace{3mm}
  \includegraphics[width=0.3\hsize]{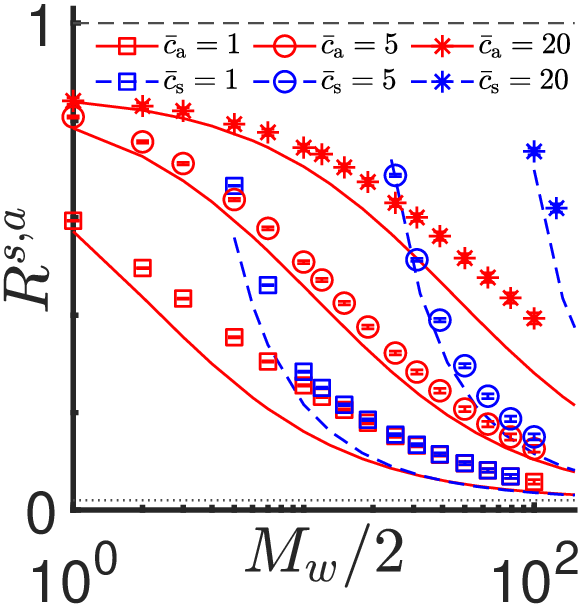}}

 \centerline{(a) \hspace{.3\hsize} \small (b) \hspace{.3\hsize} (c)}

\caption{(a) Mean photon numbers $ \langle n_{\rm s}\rangle^{\rm
a} $ [$ \langle n_{\rm i}\rangle^{\rm s} $] and (b) Fano factors
  $F_{\rm s}^{\rm a}$ [$F_{\rm i}^{\rm s}$] of the signal [idler] beam and
  (c) noise-reduction parameter $R^{\rm a}$ [$R^{\rm s}$] for signal PATWB [PSTWB] as they depend on the number $ M_{\rm w} $
  of concatenated windows in the original TWB;  {logarithmic scale is used for $ M_{\rm w}/2 $} and $ \langle n_{\rm s} \rangle^{\rm a} $, $ \langle n_{\rm i} \rangle^{\rm s} $ in (a).
  PA [PS] states are shown by red smooth [blue dashed] curves for 1 ($\scriptscriptstyle{\square}$), 5 ($\circ$), and 20 ($\ast$) added [subtracted] photons.
  Solid [dashed] curves originate in the Gaussian model while experimental data are plotted as isolated symbols with error bars (smaller than the symbols).
  Dotted lines give the values of the original TWB.
  The quantum-classical border ($ F=R = 1 $) is shown by dashed black horizontal lines in (b,c).}
\label{fig2}
\end{figure}

\emph{Model and statistical properties ---} The 1D and 2D
experimental histograms $ f $ are related to the corresponding
photon-number distributions $ p $ by the following linear
relations
\begin{eqnarray}    
 f(c_{\rm s},c_{\rm i}) &=& \sum_{n_{\rm s}=0}^{\infty}
  T(c_{\rm s},n_{\rm s}) \sum_{n_{\rm i}=0}^{\infty}
  T(c_{\rm i},n_{\rm i}) p(n_{\rm s},n_{\rm i}),
\label{1} \\
 f_{j}(c_{\rm s}) &=& \sum_{n_{\rm s}=0}^{\infty}
   T(c_{\rm s},n_{\rm s}) p_{j}(n_{\rm s}).
\label{2}
\end{eqnarray}
The detection matrix $ T $ gives the probability of having $ c $
photocounts in $ M_{\rm w} $ detection windows being illuminated
by $ n $ photons, having detection efficiency $\eta$, and mean
dark count rate $d$ \cite{Saleh1978,PerinaJr2012}:
\begin{eqnarray}     
 T(c,n;\eta,M_{\rm w},d) &=& \left( \begin{array}{c} M_{\rm w} \cr c \end{array} \right)
  (1-d)^{M_{\rm w}} (1-\eta)^{n} (-1)^{c}  \nonumber \\
  & & \hspace{-10mm} \times \sum_{l=0}^{c}
  \left( \begin{array}{c} c \cr l \end{array} \right) \frac{(-1)^l}{(1-d)^l}
  \left( 1 + \frac{l}{M_{\rm w}} \frac{\eta}{1-\eta}
   \right)^{n}.
\label{3}
\end{eqnarray}
For our experimental data, we have $\eta=0.266\pm0.005$ and
$d=6.6\times 10^{-3}$ for both beams (photocount-data
symmetrization). The photon-number distributions $ p(n_{\rm
s},n_{\rm i}) $ and $ p(n_{\rm s}) $ characterizing the analyzed
states are obtained by the inversion of the relations in
Eqs.~(\ref{1}) and (\ref{2}) accomplished by the
maximum-likelihood reconstruction method
\cite{Dempster1977,PerinaJr2012}.

According to the model presented in \cite{PerinaJr2013a}, the
photon-number distribution $ p(n_{\rm s},n_{\rm i}; M_{\rm w}) $
of a TWB detected in $ M_{\rm w} $ symmetrized double windows is
obtained as the twofold convolution of the Mandel-Rice
distributions $ p^{\rm MR}(n;M,B)= [\Gamma(n+M)/ (n! \Gamma(M))]
[B^n / (1+B)^{n+M}] $, using the $ \Gamma
$-function~\cite{Perina1991}, that characterize the photon-pair
($B_{\rm p}=0.159 $, $ M_{\rm p} = 5 M_{\rm w}/2 $), noise signal
($B_{\rm s}=0.003 $, $ M_{\rm s} = 5 M_{ w}/2 $), and noise idler
($B_{\rm i}=B_{\rm s}$, $ M_{\rm i} = M_{\rm s} $) TWB components:
\begin{eqnarray}   
 p(n_{\rm s},n_{\rm i}; M_{\rm w}) &=&  \sum_{n=0}^{\min(n_{\rm s}, n_{\rm i})}
  p^{\rm MR}(n_{\rm s}-n;M_{\rm s},B_{\rm s})
 p^{\rm MR}(n_{\rm i}-n;M_{\rm i},B_{\rm i}) p^{\rm MR}(n;M_{\rm p},B_{\rm
 p}).
\label{4}
\end{eqnarray}
We note that 5 independent modes in each of the beam components in
one symmetrized double window were identified in detailed analysis
\cite{PerinaJr2012a,PerinaJr2021b}. The mean photon numbers of
TWBs with photon-number distributions $ p(n_{\rm s},n_{\rm i};
M_{\rm w}) $ arising in $ M_{\rm w} $ concatenated detection
windows are plotted by the black dashed curve in
Fig.~\ref{fig2}(a). Using the theoretical methods of
\cite{Mandel1995,Thapliyal2024a} we obtain the photon-number distributions of
all groups of the states mentioned above (for details, see Sec.~III of SM).
We subtract 1, 5, and 20
photocounts detected in turn in $\overline{M}_{\rm s}=4 $, 22, and
90 detection windows to effectively model a beam splitter with the
transmissivities $ T_{\rm s} =1-\overline{M}_{\rm s}/M_{\rm w}$.
To allow for direct comparison of PS and PA, we add the signal
photons from the auxiliary TWBs obtained by post-selection
conditioned on the detection of 1, 5, and 20 idler photocounts in
in turn $\overline{M}_{\rm a}=4 $, 22, and 90 detection windows.
The probabilities of success in these PA and PS are given in
Tab.~\ref{tab1} comparing the experimental probabilities with
their theoretical predictions. To assess the states'
nonclassicality and quantum correlations, we introduce two simple
quantities, the Fano factor $ F $ and noise-reduction parameter $
R $:
\begin{equation}  
 F =  \langle (\Delta n)^2\rangle / \langle n\rangle, \quad R = \langle [\Delta (n_{\rm s} - n_{\rm i}) ]^2\rangle
    / (\langle n_{\rm s}\rangle + \langle n_{\rm i} \rangle );
\label{5}
\end{equation}
$ \Delta n = n - \langle n \rangle $. We note that $F<1$ ($R<1$)
identifies nonclassical (quantum-correlated) states.

\begin{figure}[t]   
 \centerline{\includegraphics[width=0.4\hsize]{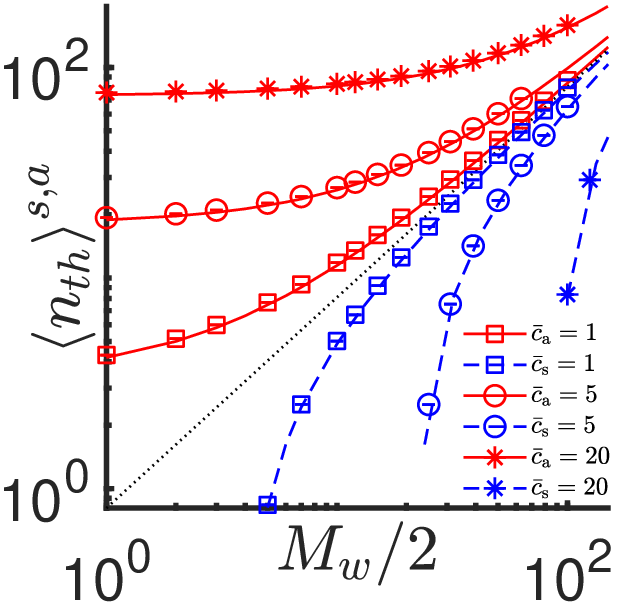}
  \hspace{6mm}\includegraphics[width=0.4\hsize]{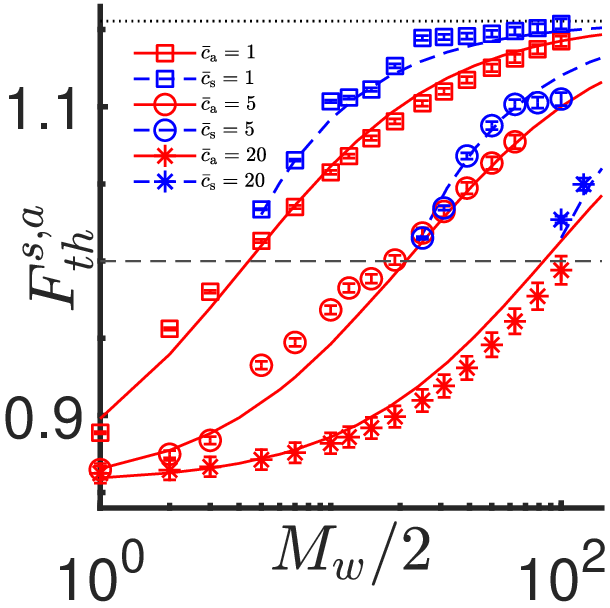}
 }
 \centerline{ (a)  \hspace{.15\hsize}  \small Original thermal states \hspace{.15\hsize} (b) }
 \vspace{2mm}
  \centerline{ \includegraphics[width=0.4\hsize]{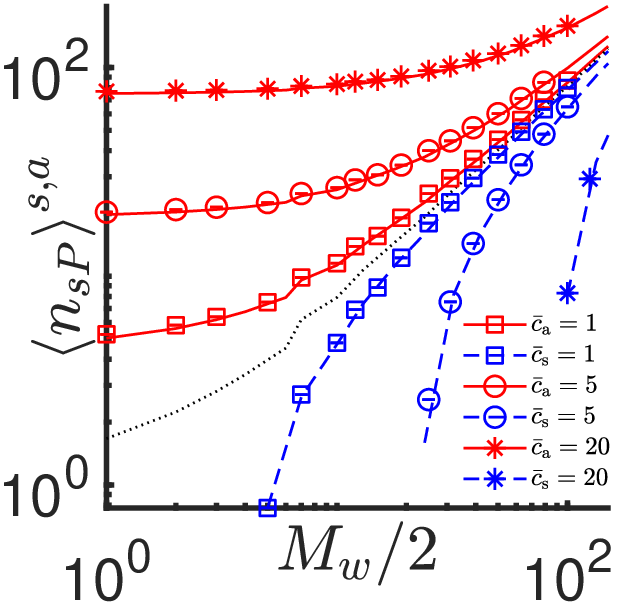}
    \hspace{6mm} \includegraphics[width=0.4\hsize]{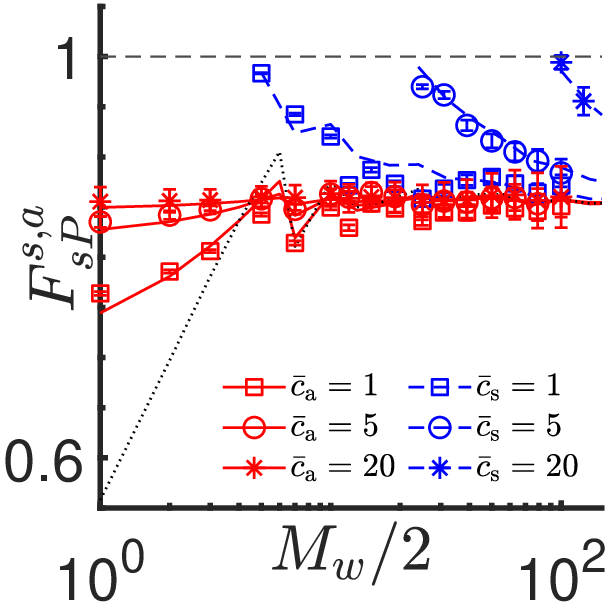}}
 \centerline{(c)  \hspace{.1\hsize} \small Original sub-Poissonian states \hspace{.1\hsize}  (d)}

 \caption{(a) [(c)] Mean photon number $ \langle n\rangle_{\rm th}^{\rm s,a} $ [$ \langle n\rangle_{\rm sP}^{\rm s,a} $] and
  (b) [(d)] Fano factor $ F_{\rm th}^{\rm s,a}$ [$ F_{\rm sP}^{\rm s,a}$] in PSTS and PATS [PSSPS and PASPS] as they depend
   on the number of concatenated windows $ M_{\rm w} $ in the original TS [SPS];  {logarithmic scale is used for $ M_{\rm w}/2 $}, $ \langle n_{\rm th} \rangle^{\rm s,a} $ in (a) and $ \langle n_{\rm sP} \rangle^{\rm s,a} $ in (c).
  Dotted curves correspond to the original TS and SPS. More details are given in the caption to Fig.~\ref{fig2}.
  }
\label{fig3}
\end{figure}

\emph{Experimental data analysis ---} First, we address PA and PS
in TWBs. We perform PA to the signal beam of the TWB which leads
to the nonclassicality of the signal beam. To provide fair
comparison between PA and PS, we subtract the same number of
photocounts from the signal beam and analyze the nonclassicality
induced in the idler beam, whose photon numbers are tightly
correlated with those of the signal beam. Whereas PA in the signal
beam linearly increases the signal mean photon number, PS from the
signal beam keeps the mean number of idler photons unchanged [see
Fig.~\ref{fig2} (a)]. Both PA and PS induce the nonclassicality at
the expense of quantum correlations between the beams, as
quantified by the noise-reduction parameter $ R $ in
Fig.~\ref{fig2} (c). PS is superior over PA under comparable
conditions, as documented in Fig.~\ref{fig2} (b) [compare the red
and blue symbols and curves]. However, PS pays the price for
it--- it disturbs more the quantum correlations, as quantified by
$ R $ in Fig.~\ref{fig2} (c). Smaller Fano factors $F_{\rm i}^{\rm
s}$ in PS compared to the Fano factors $F_{\rm s}^{\rm a}$ in PA
are attributed to smaller mean photon numbers $ \langle n_{\rm
i}\rangle^{\rm s} $ in PS compared to the mean photon numbers $
\langle n_{\rm s}\rangle^{\rm a} $ in PA. We note that the schemes
for PA and PS in compound TWBs drawn in terms of the basic
building blocks [see Figs.~\ref{fig1} (b,c)] show one-to-one
mapping between PA and PS. The effects of PA and PS become
comparable as the intensity of the original TWB increases, i.e.
when the numbers of added or subtracted photons are smaller or
comparable to the mean number of photon pairs in the TWB. Once the
mean photon-pair number in the TWB is considerably larger than the
number of added or subtracted photons, PA and PS modify the
original TWB negligibly. Comparing the curves in Figs.~\ref{fig2}
(b,c) we conclude in general that the greater the number of added
or subtracted photons is, the stronger the changes in the original
TWB are, the more nonclassical the marginal beams are, and the
weaker the remaining quantum correlations in the TWB are.

Analyzing the signal beams of the above TWBs, we demonstrate the
properties of PA and PS in multi-mode TSs. According to the curves
in Fig.~\ref{fig3} (a,b) PA increases the mean photon numbers $
\langle n\rangle_{\rm th}^{\rm a} $ and simultaneously generates
the nonclassicality: The greater the number of added photons is,
the stronger and the more nonclassical the beam is. Contrary to
this, PS decreases the beam intensity and, though it lowers the
values of the Fano factor $ F_{\rm th}^{\rm s}$, it cannot provide
a nonclassical state.

Post-selecting by detection of a fixed number of idler photocounts
we obtain SPSs in the signal beam. Similarly as for TSs, PA [PS]
increases [decreases] the mean signal photon number $ \langle
n\rangle_{\rm sP}^{\rm a} $ [$ \langle n\rangle_{\rm sP}^{\rm s}
$]. However, none of these processes significantly enhances the
signal-beam nonclassicality naturally present in SPSs, as
illustrated in Fig.~\ref{fig3} (d). We note that the mean photon
number $ \langle n\rangle_{\rm sP} $ of the original SPS increases
with the number $M_{\rm w}$ of concatenated detection windows, as
the number $c_{\rm i}$ of the used post-selecting idler
photocounts, appropriate to the most probable post-selection
event, systematically increases with the increasing $M_{\rm w}$.

PA and PS in the above beams share one common property: Once the
numbers of added or subtracted photons are considerably lower than
the mean number of photons or photon-pairs in a beam, the effects
induced by PA and PS are negligible.

\begin{figure}[t]   
 \centerline{\includegraphics[width=0.4\hsize]{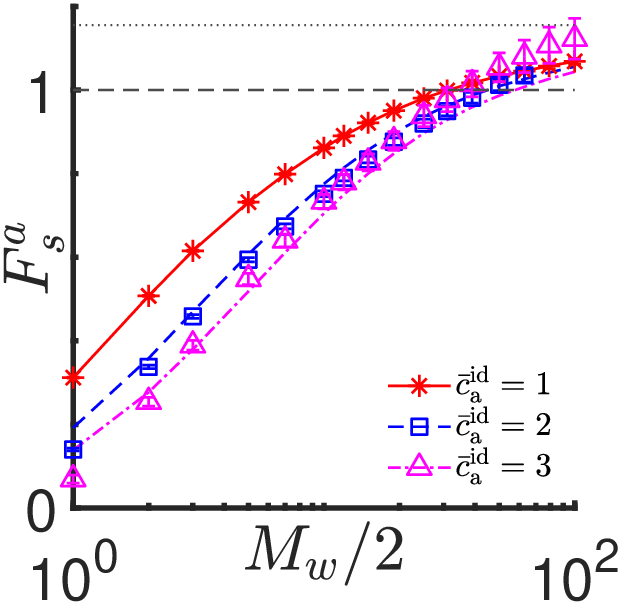}
   \hspace{6mm} \includegraphics[width=0.4\hsize]{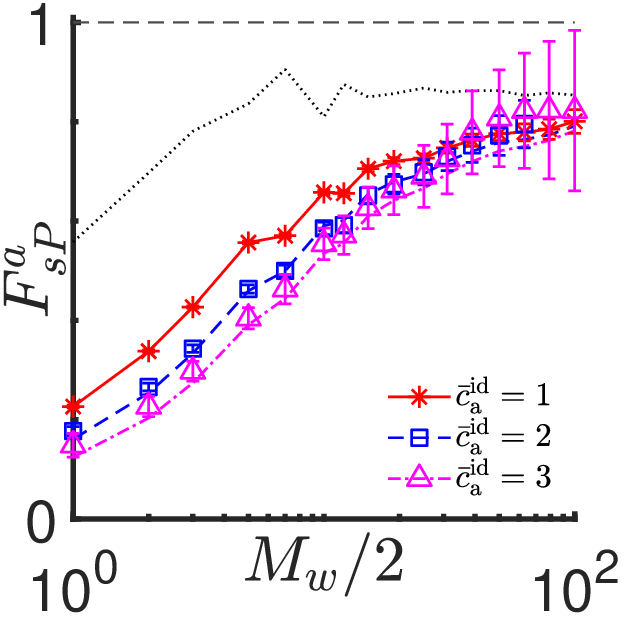}

 }
 \centerline{ (a)  \hspace{.3\hsize}   (b) }

 \caption{(a) [(b)]
  Fano factor $ F_{\rm th}^{\rm a} \equiv F_{\rm s}^{\rm a}$ [$ F_{\rm sP}^{\rm a}$] for ideal-PA in TS [SPS] as it
  depends on the number $ M_{\rm w} $ of concatenated windows for 1 (red $\ast$), 2 (blue $ \scriptscriptstyle{\square}$),
  and 3 (magenta $\Delta$) ideally-added photons;  {logarithmic scale is used for $ M_{\rm w}/2 $}.
  Curves originate in the Gaussian model while experimental data are plotted as isolated symbols with error bars.
  The black dotted curves characterize the original TS and SPS.
  The quantum-classical border $ F = 1 $ is shown by dashed black horizontal line.}
\label{fig4}
\end{figure}

The technique of compound beams also allows to mimic the operation
of an ideal photon-number-resolving detector in post-selection
once temporal correlations of photons in weak TWBs--- the building
blocks of compound beams ---are exploited. In this case, we can
speak about ideal PA. As the experimental probability of ideal PA
is considerably smaller than that of the usual PA provided in
Tab.~\ref{tab1}, we only analyze the states with 1, 2, and 3
photons ideally added, whose probabilities of generation are in
turn 8.10\%, 1.05\%, and 0.18\% (relative error 1\%). Ideal
addition of photons naturally results in greater nonclassicality
of the ideally PA states compared to their real PA states, as
apparent from direct comparison of the curves for TSs and SPSs
drawn in Fig.~\ref{fig3}(b) vs. \ref{fig4}(a) and
Fig.~\ref{fig3}(d) vs. \ref{fig4}(b). In contrast to the operation
of PA in SPSs, ideal PA in SPSs allows to considerably increase
the beam nonclassicality expressed via the beam Fano factor $
F_{\rm sP}^{\rm a}$ in \ref{fig4}(b). We note that the application
of ideal PA reveals the true potential of this non-Gaussian
operation and it may potentially open new avenues for developing
quantum technologies.

\begin{figure}[t]   
\centerline{ \includegraphics[width=0.4\hsize]{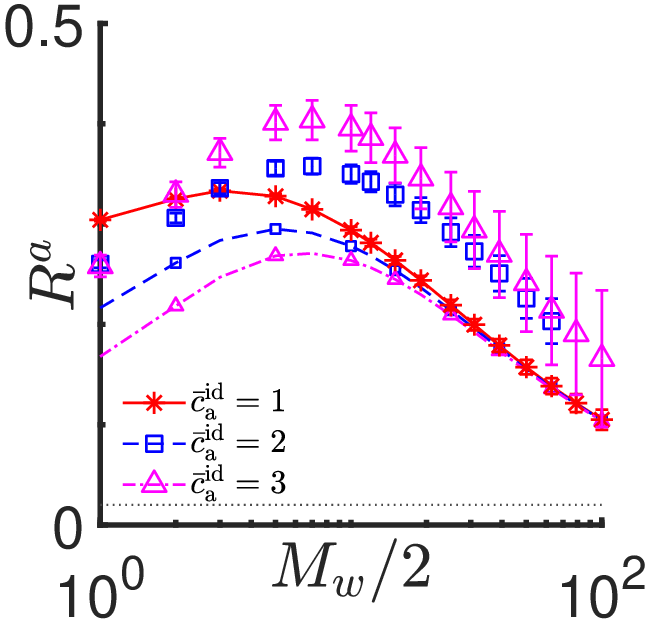}
  \hspace{6mm} \includegraphics[width=0.4\hsize]{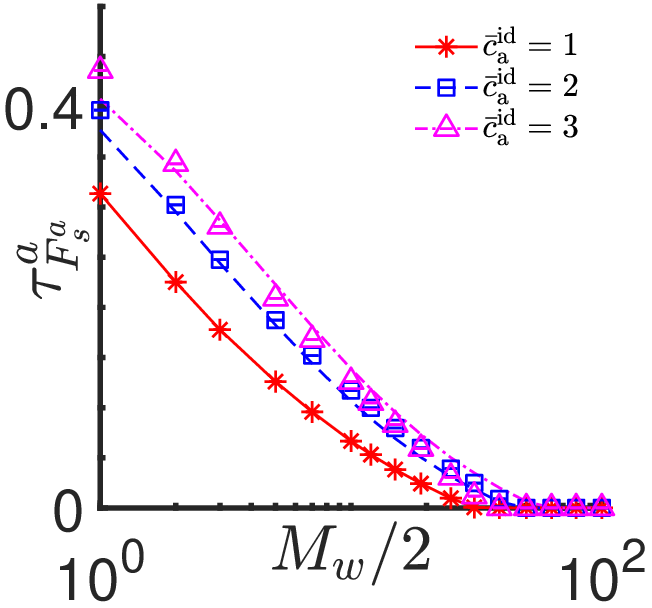}}
 \centerline{ (a)  \hspace{.25\hsize}   \hspace{.25\hsize} (b) }
  \vspace{2mm}
  \centerline{ 
    \includegraphics[width=0.4\hsize]{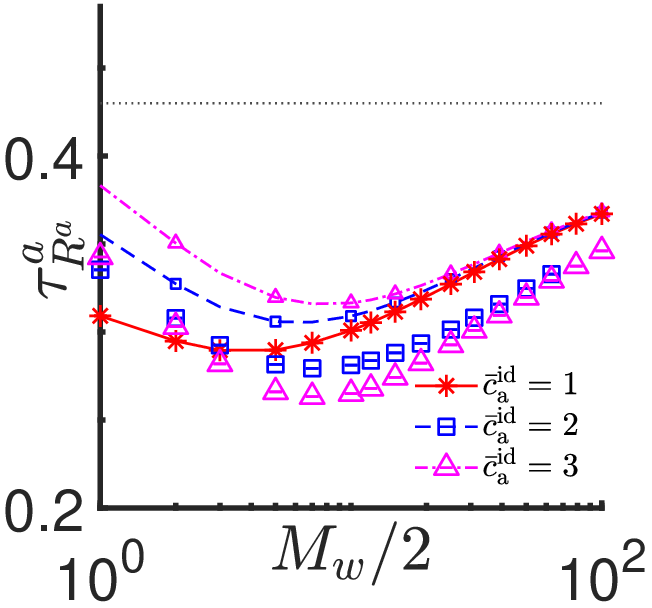}
      \hspace{6mm} \includegraphics[width=0.4\hsize]{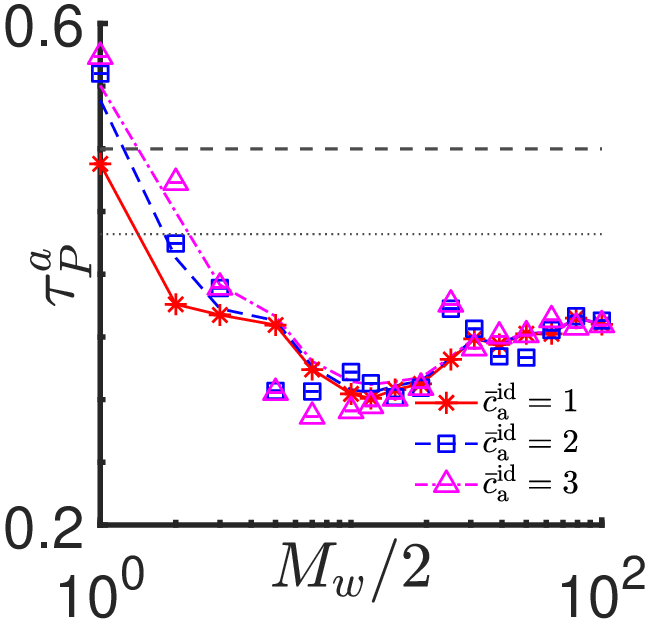}}
 \centerline{ (c)  \hspace{.25\hsize}   \hspace{.25\hsize} (d) }

 \caption{
   (a) Noise-reduction parameter $ R^{\rm a} $,
   and nonclassicality depths (b) $\tau^{\rm a}_{F^{\rm a}_{\rm s}}$, (c)
   $\tau^{\rm a}_{R^{\rm a}}$, and (d) $\tau^{\rm a}_{P} $  as they depend on the number $ M_{\rm w} $
   of concatenated windows for 1 (red $\ast$), 2 (blue $ \scriptscriptstyle{\square}$), and 3 (magenta
   $\Delta$) photons ideally added into TWBs;  {logarithmic scale is used for $ M_{\rm w}/2 $}.
   Curves  {[in (a,c) marked by symbols]} originate in the Gaussian model while experimental data are plotted as isolated symbols with error bars.
   The black dotted lines characterize the original TWBs.
   The Gaussian border $ \tau = 0.5 $ in (d) is indicated by dashed black horizontal line.
   In (b,c,d) relative errors are smaller than 10\% for smaller $ M_{\rm w} $
   and 20\% for larger $ M_{\rm w} $.
  }
\label{fig5}
\end{figure}

\emph{More advanced experimental data analysis ---}   Fano factor
$ F $ and noise-reduction parameter $ R $ are commonly used to
evidence and somehow quantify the nonclassicality (quantum
correlations) of the beams because of their simple determination.
Nevertheless, they are not exact quantifiers of the
nonclassicality. According to its definition
\cite{Glauber1963,Sudarshan1963} the nonclassicality depth $ \tau
$ introduced by Lee~\cite{Lee1991} is considered as the most
natural and also convenient nonclassicality quantifier. The value
of $ \tau $ is directly derived from the threshold value of the
field-operator-ordering parameter $ s_{\rm th} $ at which a field
loses its nonclassical features. Physically, the nonclassicality
depth $ \tau $ gives the amount of thermal noise needed to conceal
the field's nonclassical features. It can directly be determined
from its definition though this may be rather involved as the
field intensity quasi-distributions \cite{Perina1991} have to be
reconstructed from the corresponding photocount moments. To
demonstrate the capability of Fano factor $ F $ and
noise-reduction parameter $ R $ in rough quantification of the
nonclassicality, we convert them to the corresponding
nonclassicality witnesses \cite{PerinaJr2017a} and assign to them
the corresponding nonclassicality depths. The nonclassicality
depths obtained this way give the lower bounds for the actual
nonclassicality depths. Relying on the relations between the
photon-number and intensity (normally-ordered photon-number)
moments, as described by the Stirling numbers
\cite{Gradshtein2000}, the quantum-classical borders $ F=1 $ and $
R=1 $ are identified by the nonclassicality witnesses $ I_{F} $
and $ I_R $, respectively:
\begin{eqnarray}   
 I_{F} & = &  \langle [\Delta W_{\rm s}]^2 \rangle <0,
 \label{6} \\
 I_{R} & = &  \langle [ \Delta( W_{\rm s} - W_{\rm i})]^2 \rangle
 <0.
 \label{7}
\end{eqnarray}

These quantifiers are mutually compared in Fig.~\ref{fig5} where
ideal PA of small numbers of photons into TWBs is analyzed.
According to the curves in Figs.~\ref{fig4}(a) and \ref{fig5}(b)
the signal-beam Fano factor $ F^{\rm a}_{\rm s} $ [more precisely
$ 1-F^{\rm a}_{\rm s} $] considered as a function of the number $
M_{\rm w} $ of concatenated windows and number $ \bar{c}_{\rm a} $
of added photons shows the same dependencies as the corresponding
nonclassicality depth $ \tau_{F^{\rm a}_{\rm s}} $. Similarly, the
noise-reduction parameter $ R^a $ [more precisely $ 1-R^a $] and
the corresponding nonclassicality depth $ \tau_{R^a} $ drawn in
Figs.~\ref{fig5}(a,c) synchronously identify qualitative change of
the behavior of nonclassicality of PATWBs with the increasing
number $ \bar{c}_{\rm a} $ of ideally added photons when moving
from weak to stronger PATWBs. This behavior is confirmed by the
actual values of the nonclassicality depths $ \tau_{P} $ derived
from the corresponding intensity quasi-distributions and plotted
in Figs.~\ref{fig5}(d). We note that whereas the actual values of
$ \tau_{P} $ are considerably greater than the values of $
\tau_{R^a} $ for small TWB intensities, they are comparable for
greater TWB intensities. We also note that the values of $
\tau_{P} $ exceed 0.5 for PATWBs with low intensities which
certifies their non-Gaussian character.

\section{Conclusions}
Compound beams with their versatile structures have been suggested
for direct experimental comparison of different quantum operations
used in the generation of highly nonclassical states occurring in
various quantum-information protocols. As an example, the
properties of the states obtained by multiple photon addition and
photon subtraction in twin beams, thermal states, and
sub-Poissonian states were mutually compared. Both quantum
operations were found as efficient tools for inducing or enhancing
the state nonclassicality. Whereas photon addition outperforms
photon subtraction in thermal and sub-Poissonian states, it looses
when applied to twin beams. The technique of compound beams
applied to the fields with mean photon numbers per mode much
less than one has been shown to be able to study side-by-side the
beams with intensities differing in orders of magnitude as well as
to mimic post-selection by an ideal photon-number-resolving
detector. The structures of the analyzed compound beams as
composed of their elementary experimental building blocks can be
generalized to incorporate the operations that implement suitable
quantum information protocols realized with the compound beams.
This gives the technique a great potential not only for optimizing
the properties of different kinds of nonclassical light sources
but also for direct implementation of various quantum
communication, metrology and information protocols. The schemes
based on the temporal compound beams can alternatively be mapped
onto real experimental setups using naturally multi-mode beams
being detected by spatially-resolved photon-number-resolving
detectors.

%
%


\funding{K.T. acknowledges funding from the European
Union's Horizon Europe research and innovation programme under the
Marie Sklodowska-Curie grant agreement No. 101126636. K.T., J.P., and A.\v{C}. 
acknowledge support by the project OP JAC
CZ.02.01.01/00/22\_008/0004596 of the Ministry of Education,
Youth, and Sports of the Czech Republic.}


\data{Data underlying the results presented in this paper are available in Ref.~\cite{Data}.}


\bibliographystyle{iopart-num}
\bibliography{thapliyal}

\end{document}


\title{Compound beams for direct experimental comparison of quantum operations: supplementary material}

\author{Kishore Thapliyal$^{1,2,*}$\orcid{0000-0002-4477-6041}, Jan Pe\v{r}ina Jr.$^{2,3}$, Pavel Pavl\'\i\v{c}ek$^3$,  and Anton\' \i n \v{C}ernoch$^{3}$}

\affil{$^1$Department of Physics,
University of Oslo, 0316 Oslo, Norway}

\affil{$^2$Joint Laboratory of
Optics, Faculty of Science, Palack\'{y} University, Czech Republic, 17. listopadu 12, 779~00 Olomouc, Czech Republic}

\affil{$^3$Joint Laboratory of Optics of Palack\'{y} University
and Institute of Physics of the Czech Academy of Sciences,
Institute of Physics of the Czech Academy of Sciences, 17.
listopadu 1154/50a, 779 00 Olomouc, Czech Republic}

\affil{$^*$Author to whom any correspondence should be addressed.}

\email{kishort@fys.uio.no}

\begin{abstract}
In this Supplemental Material, we present the approach for
constructing experimental photocount histograms for the analyzed
states. Also the determination of nonclassicality depth using both
suitable nonclassicality criteria and directly quasi-distributions
of integrated intensities of photon-added and -subtracted twin
beams is discussed in detail. The theoretical model for the analyzed states is presented in Sec.~III.
\end{abstract}

\section{Experimental reconstruction of frequency histograms}

We give details of the method of obtaining the experimental
frequency histograms for the states discussed in the main text:
twin beams (TWBs), thermal states (TSs), sub-Poissonian states
(SPSs), photons-added twin beams (PATWBs), photons-subtracted twin
beams (PSTWBs), photons-added thermal states (PATSs),
photons-subtracted thermal states (PSTSs), photons-added
sub-Poissonian states (PASPSs), photons-subtracted sub-Poissonian
states (PSSPSs). The schemes for constructing the frequency
histograms of different groups of the states are provided in
Fig.~1 of the main text. Each event in the frequency histogram $
f(c_{\rm s},c_{\rm i}; M_{\rm w}) $ of a TWB with $ c_{\rm s} $
signal and $ c_{\rm i} $ idler photocounts is obtained by
concatenating neighbour $M_{\rm w}$ detection windows in two
synchronized (signal and idler) channels whose individual signal
($ \tilde{c}_{\rm s}$) and idler ($ \tilde{c}_{\rm i}$) photocount
numbers characterize detections of weak TWBs and equal 0 or 1.
Entry to the histogram $ f(c_{\rm s},c_{\rm i}; M_{\rm w}) $
starting at position $ j $ in these channels is characterized by
the photocount numbers $ c_{\rm s} $ and $ c_{\rm i} $ given as:
\begin{eqnarray}  
 c_{\rm s}(j) &=& \sum_{k=1}^{M_{\rm w}/2} \left(\tilde{c}_{{\rm s}_{j+2k-2}} + \tilde{c}_{{\rm i}_{j+2k-1}}\right),
 \label{S1} \\
 c_{\rm i}(j) &=& \sum_{k=1}^{M_{\rm w}/2} \left(\tilde{c}_{{\rm i}_{j+2k-2}} + \tilde{c}_{{\rm
 s}_{j+2k-1}}\right).
\label{S2}
\end{eqnarray}
For large numbers $\approx M_{\rm m} $ of experimental
realizations of weak TWBs, we have close to $ M_{\rm m} $ entries
into the histogram $ f(c_{\rm s},c_{\rm i}; M_{\rm w}) $. We note
that we use multiple times the photocounts of weak TWBs. We also note
that the formulas (\ref{S1}) and (\ref{S2}) assure symmetrization
of the photocount histogram with respect to the photocount data
embedded in the signal and idler detection channels.

The frequency histograms of marginal TS $ f_{\rm th}(c_{\rm s};
M_{\rm w}) $ and SPS $ f_{\rm sP}(c_{\rm s};c_{\rm i}, M_{\rm w})
$ post-selected by detecting $ c_{\rm i} $ idler photocounts are
derived from that of the TWB written in Eqs.~(\ref{S1}) and
(\ref{S2}):
\begin{eqnarray}   
 f_{\rm th}(c_{\rm s}; M_{\rm w})&=& \sum_{c_{\rm i}=0}^{\infty} f(c_{\rm
   s},c_{\rm i}; M_{\rm w}), \label{S3} \\
 f_{\rm sP}(c_{\rm s};c_{\rm i}, M_{\rm w}) &=& f(c_{\rm s},c_{\rm i}; M_{\rm w})
  / \sum_{c_{\rm s}=0}^{\infty} f(c_{\rm
   s},c_{\rm i}; M_{\rm w}) .
\label{S4}
\end{eqnarray}
During SPS generation with fixed number of $M_{\rm w}$ detection
windows, we select the conditioning idler photocount number
$c_{\rm i}$ that has the maximal success probability:
\begin{eqnarray}  
  p_{\rm suc}(c_{\rm i}; M_{\rm w})= \sum_{c_{\rm s}=0}^{\infty} f(c_{\rm s},c_{\rm i}; M_{\rm w})/\sum_{c_{\rm s},c_{\rm i}=0}^{\infty} f(c_{\rm s},c_{\rm i}; M_{\rm w}).
\label{S5}
\end{eqnarray}

\begin{figure}[t]   
\centerline{\includegraphics[width=0.47\hsize]{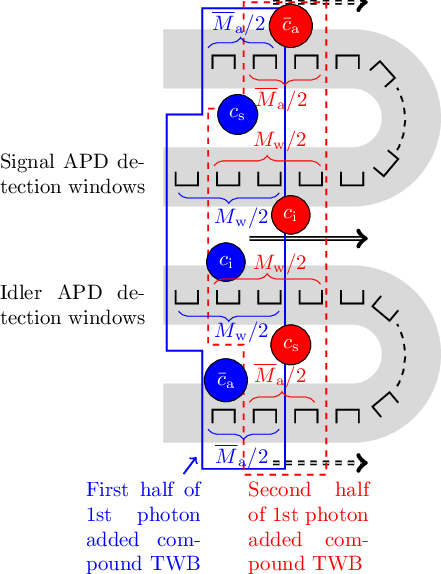} 
\includegraphics[width=0.47\hsize]{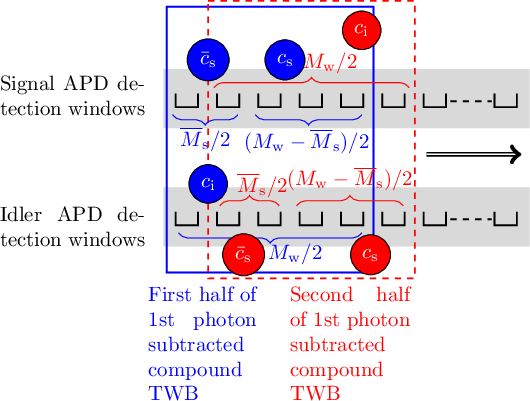} 
} 
 \centerline{ (a)  \hspace{.45\hsize}   (b)}
 \caption{Schematic diagrams elucidating processing the photocount data in the signal and the idler experimental photocount channels when constructing the photocount histograms of (a) PATWBs and (b) PSTWBs.
 To compensate for different detection parameters of the signal and idler APDs, individual realizations of PATWBs
 and PSTWBs are constructed from the basic building blocks containing the neighbor two detection windows in each beam such that the role of the
 signal and the idler beams is exchanged in the second
 detection window. This exchange is indicated by the red color. Whereas the photocount data for PATWB are taken synchronously from both ends of the channels (a), the data for PSTWB are read only from the beginning of the channels (b). The data read from the end of the channels are used to construct the auxiliary TWB from which the states to be added into the TWB are obtained by post-selection. Subsequent realizations of PATWBs and PSTWBs 
 are obtained by moving along the channels in the directions indicated by the arrows (TWB: $\longrightarrow$, state for PA: $\dashrightarrow$).}
\label{figS1}
\end{figure}

For photons addition (PA), we use post-selection from an auxiliary
TWB$_{\rm a}$ whose signal ($ c_{\rm s, a} $) and idler ($ c_{\rm i, a}
$) photocounts detected in $ \overline{M}_{\rm a} $ neighbor
detection windows are taken from the end of the signal and idler
detection channels:
\begin{eqnarray}  
 c_{\rm s,a}(j)  =
 \sum_{k=1}^{\overline{M}_{\rm a}/2} \left(\tilde{c}_{{\rm s}_{M_{\rm m}-(j+2k-3)}} + \tilde{c}_{{\rm i}_{M_{\rm m}-(j+2k-2)}}\right),
 \nonumber \\
 c_{\rm i,a}(j)  =
 \sum_{k=1}^{\overline{M}_{\rm a}/2} \left(\tilde{c}_{{\rm i}_{M_{\rm m}-(j+2k-3)}} + \tilde{c}_{{\rm
   s}_{M_{\rm m}-(j+2k-2)}}\right).
 \label{S6}
\end{eqnarray}
The index $ j $ in Eq.~(\ref{S6}) refers to the position in the
signal and idler detection channels set when generating the
entries to the TWB histogram $ f(c_{\rm s},c_{\rm i}; M_{\rm w}) $ along Eqs.~(\ref{S1}) and (\ref{S2}). A schematic diagram for processing experimental data for PATWB is given in Fig.~\ref{figS1} (a).

Using both TWB and auxiliary TWB, entry to the frequency histogram
$f^{\rm a}(c_{\rm s},c_{\rm i}; \bar{c}_{\rm a},M_{\rm w})$ of a
PATWB with $ \bar{c}_{\rm a} = c_{\rm i,a}(j)$ photocounts
added, defined in Eq.~(\ref{S6}), is characterized by the following photocount numbers:
\begin{eqnarray}  
 c_{\rm s}^{\rm a}(j)  &=& \sum_{k=1}^{M_{\rm w}/2} (\tilde{c}_{{\rm s}_{j+2k-2}} + \tilde{c}_{{\rm
   i}_{j+2k-1}})
 + \sum_{k=1}^{\overline{M}_{\rm a}/2} (\tilde{c}_{{\rm s}_{M_{\rm m}-(j+2k-3)}} + \tilde{c}_{{\rm i}_{M_{\rm m}-(j+2k-2)}}).\nonumber \\
 c_{\rm i}^{\rm a}(j) &=& \sum_{k=1}^{M_{\rm w}/2} \left(\tilde{c}_{{\rm i}_{j+2k-2}} + \tilde{c}_{{\rm
 s}_{j+2k-1}}\right).
 \label{S7}
\end{eqnarray}
For fixed number $ \bar{c}_{\rm a} $ of added photocounts, the
number $\overline{M}_{\rm a}$ of concatenated detection windows is
set such that the success probability $ p_{\rm succ} $ determined
in analogy to Eq.~(\ref{S5}) is maximal.

The frequency histograms of PATS $ f_{\rm th}^{\rm a}(c_{\rm
s};\bar{c}_{\rm a},M_{\rm w}) $ and PASPS $ f_{\rm sP}^{\rm
a}(c_{\rm s};c_{\rm i},\bar{c}_{\rm a},M_{\rm w}) $ post-selected
by detecting $ c_{\rm i} $ idler photocounts are determined from
that of the PATWB written in Eq.~(\ref{S7}):
\begin{eqnarray}   
 f_{\rm th}^{\rm a}(c_{\rm s};\bar{c}_{\rm a}, M_{\rm w})&=& \sum_{c_{\rm i}=0}^{\infty} f^{\rm a}(c_{\rm
   s},c_{\rm i};\bar{c}_{\rm a}, M_{\rm w}), \label{S8} \\
 f_{\rm sP}^{\rm a}(c_{\rm s};c_{\rm i},\bar{c}_{\rm a}, M_{\rm w}) &=& f^{\rm a}(c_{\rm s},c_{\rm i};\bar{c}_{\rm a}, M_{\rm w})
 / \sum_{c_{\rm s}, c_{\rm i} =0}^{\infty} f^{\rm a}(c_{\rm
   s},c_{\rm i};\bar{c}_{\rm a}, M_{\rm w}) .
\label{S9}
\end{eqnarray}

When implementing photon subtraction (PS) in the signal beam, the
$ M_{\rm w} $ neighbor signal detection windows are divided into
two groups with $\overline{M}_{\rm s}$ and $ M_{\rm w}
-\overline{M}_{\rm s}$ detection windows. Whereas the $ M_{\rm w}
-\overline{M}_{\rm s}$ detection windows remain in the signal
beam, the $\overline{M}_{\rm s}$ detection windows represent the
beam reflected on the (virtual) beam spitter. The corresponding
photocount numbers $ c_{\rm s}(j) $ and $ c^{\rm s}_{\rm s}(j) $
belonging to the entry to the original histogram $ f(c_{\rm
s},c_{\rm i}; M_{\rm w}) $ of a TWB starting at position $ j $ are
given as:
\begin{eqnarray}  
 c_{\rm s}(j)  &=& \sum_{k=\overline{M}_{\rm s}/2+1}^{M_{\rm w}/2} \left(\tilde{c}_{{\rm s}_{j+2k-2}} + \tilde{c}_{{\rm i}_{j+2k-1}}\right), \label{S10} \\
 c_{\rm s}^{\rm s}(j) & = &
 \sum_{k=1}^{\overline{M}_{\rm s}/2} \left(\tilde{c}_{{\rm s}_{j+2k-2}} + \tilde{c}_{{\rm
 i}_{j+2k-1}}\right).
 \label{S11}
\end{eqnarray}
The corresponding entry to the frequency histogram $f^{\rm
s}(c_{\rm s},c_{\rm i}; \bar{c}_{\rm s},M_{\rm w})$ of a PSTWB
with $ \bar{c}_{\rm s} \equiv c_{\rm s}^{\rm s}(j) $ subtracted
signal photocounts is characterized by detected $ c_{\rm s}(j) $
signal photocounts and $ c_{\rm i}(j) $ idler photocounts given in
Eqs.~(\ref{S10}) and (\ref{S2}), respectively. Figure~\ref{figS1} (b) shows a schematic diagram for processing experimental data for PSTWB. For fixed number
$\bar{c}_{\rm s}$ of subtracted signal photocounts, the number
$\overline{M}_{\rm s}$ of detection windows is chosen to maximize
the success probability $ p_{\rm succ} $ given  in analogy to
Eq.~(\ref{S5}). 

The frequency histograms of PSTS $ f_{\rm th}^{\rm s}(c_{\rm
s};\bar{c}_{\rm s},M_{\rm w}) $ and PSSPS $ f_{\rm sP}^{\rm
a}(c_{\rm s};c_{\rm i},\bar{c}_{\rm s},M_{\rm w}) $ post-selected
by detecting $ c_{\rm i} $ idler photocounts are then derived from
the histogram $f^{\rm s}(c_{\rm s},c_{\rm i}; \bar{c}_{\rm
s},M_{\rm w})$ of the PSTWB:
\begin{eqnarray}   
 f_{\rm th}^{\rm s}(c_{\rm s};\bar{c}_{\rm s}, M_{\rm w})&=& \sum_{c_{\rm i}=0}^{\infty} f^{\rm s}(c_{\rm
   s},c_{\rm i};\bar{c}_{\rm s}, M_{\rm w}), \label{S12} \\
 f_{\rm sP}^{\rm s}(c_{\rm s};c_{\rm i},\bar{c}_{\rm s}, M_{\rm w}) &=& f^{\rm s}(c_{\rm s},c_{\rm i};\bar{c}_{\rm s}, M_{\rm w})
 / \sum_{c_{\rm s}, c_{\rm i} =0}^{\infty} f^{\rm s}(c_{\rm
   s},c_{\rm i};\bar{c}_{\rm s}, M_{\rm w}) .
\label{S13}
\end{eqnarray}

\section{Nonclassicality depth of photon-added and -subtracted twin beams}

Nonclassicality is identified by definition by negative values of
the Glauber-Sudarshan $P$-function
\cite{Glauber1963,Sudarshan1963} or by its existence as a
generalized function. Its negativity can equivalently be evidenced
by (infinite) number of inequalities for the moments of
annihilation and creation operators \cite{Miranowicz2010}. These
inequalities are usually called nonclassicality criteria,
witnesses or identifiers.

In photon-counting experiments, only intensity moments are
obtained. Nevertheless, for specific groups of states, there exist
efficient nonclassicality criteria (for Gaussian fields, see
\cite{PerinaJr2017a}, for non-Gaussian fields, see
\cite{Thapliyal2024a,PerinaJr2024}). The nonclassicality criteria
$ I_F $ and $ I_R $ defined in Eqs.~(6) and (7) in the main text
and derived from the definitions of the Fano factor $ F $ and the
noise-reduction parameter $ R $ belong to the most efficient. The
reason is that they are based on second-order intensity moments,
and it usually holds that with the increasing order of intensity
moments the efficiency of the nonclassicality criteria drops down
\cite{PerinaJr2022}.

The nonclassicality criteria can not only identify the
nonclassicality, they can also quantify its strength. Lee
introduced the concept of nonclassicality depth $ \tau $
\cite{Lee1991} for this quantification. The nonclassicality depth
$ \tau $  is defined in terms of the threshold value $ s_{\rm th}
$ of the ordering parameter $s$ at which the nonclassicality
present in the state is concealed by the detection noise
\cite{Perina1991} [$ \tau \equiv (1-s_{\rm th})/2 $]. Whereas the
exact value of $ s_{\rm th} $ can only be determined from the form
of quasi-distribution $ P_{s} $ of field amplitudes related to a
general $ s $-ordering of field operators, it can be estimated
from below by transforming the intensity moments to a general $ s
$-ordering of field operators and using suitable nonclassicality
criteria \cite{PerinaJr2017a}. We note that a $ k $th-order
intensity (normally-ordered photon-number) moment is determined in
terms of the usual photon-number moments $ \langle n^l\rangle $
as~\cite{Perina1991}
\begin{equation}   
 \langle W^k\rangle = \sum_{l=1}^{k} S_{kl} \langle n^l\rangle
\label{S14}
\end{equation}
using the Stirling numbers $ S_{kl} $ of the first kind
\cite{Gradshtein2000}. Then, an $s$-ordered intensity moment
$\langle W^k\rangle_{\rm s} $ is obtained as~\cite{Perina1991}
\begin{equation}   
 \langle W^k\rangle_{\rm s} = \frac{k!}{\Gamma\left( k+M\right)}\left(\frac{1-s}{2}\right)^{k}
 \Big\langle L_{k}^{M-1}\left(\frac{2W}{s-1}\right)
 \Big\rangle
\label{S15}
\end{equation}
$ L_x^n $ stands for the associated Laguerre polynomials
\cite{Morse1953,Perina1991} and $ M $ denotes the number of
field's modes.

When photocount distributions are measured for multi-mode optical
fields, complete characterization of the state is achieved by
determining the corresponding quasi-distribution $ P $ of
intensities. This can be done considering a general $ s $-ordering
of field operators. For a TWB analyzed in the main text, the
following formula is applied~\cite{Perina1991}:
\begin{eqnarray} 
 P_{\rm s}(W_{\rm s},W_{\rm i})&=& \frac{4}{(1-s)^2} \exp\left(-\frac{2(W_{\rm s}+W_{\rm i})}{1-s}\right)
 \sum_{n_{\rm s},n_{\rm i} =0}^{\infty}  \frac{p(n_{\rm s},n_{\rm i})}{n_{\rm s}!\, n_{\rm i}!}
  \left(\frac{s+1}{s-1}\right)^{n_{\rm s}+n_{\rm i}}  
  \nonumber \\
 & & \hspace{-7mm} \times 
 L_{n_{\rm s}}^{M_{\rm s}-1}\left(\frac{4W_{\rm s}}{1-s^2}\right)
   L_{n_{\rm i}}^{M_{\rm i}-1}\left(\frac{4W_{\rm i}}{1-s^2}\right)
\label{S16}
\end{eqnarray}
and $ p(n_{\rm s},n_{\rm i})$ denotes the reconstructed
photon-number distribution of the TWB. Symbol $ M_{\rm s} $ ($
M_{\rm i} $) stands for the number of modes in the signal (idler)
beam. Nevertheless, the application of formula~(\ref{S16}) is
numerically demanding and it can be computed only around $ s_{\rm
th} $ and for $ s< s_{\rm th} $. This is the reason why the
nonclassicality depths $ \tau $ (lower bounds) determined for
suitable nonclassicality criteria are extraordinarily useful.

\section{Model for photon-added and photon-subtracted states generated from twin beams}

Thermal and sub-Poissonian states and twin beams, as well as their PA and PS variants, analyzed in the main text are described using specific forms of two TWBs~\cite{Thapliyal2024a}. The first 
twin beam TWB$_{\rm p} $ is used to describe the original thermal and sub-Poissonian states
and twin beams, whereas the second twin beam TWB$_{\rm a} $ serves as the source of
the fields for PA. Both TWBs are assumed in multi-mode Gaussian states with their photon-pair, noise signal and noise idler components, as described in Eq.~(4) of the main text. 

Photon-number distributions of the analyzed groups of states are derived from 
the photon-number distributions $ p^{\rm TWB}_{\rm p}(n_{\rm s},n_{\rm i}) $ and
$ p^{\rm TWB}_{\rm a}(n_{\rm s},n_{\rm i}) $ of the twin beams TWB$ _{\rm
p} $ and TWB$_{\rm a} $, respectively, as follows.

\begin{description}
 \item[TS] A multi-mode thermal state (TS) is obtained as a marginal state of TWB$_{\rm p}$:
  \begin{equation}  
   p_{\rm th}(n_{\rm s}) = \sum_{n_{\rm i}=0}^{\infty}p^{\rm TWB}_{\rm p}(n_{\rm s},n_{\rm i}).
  \label{17}
  \end{equation}

 \item[SPS] A sub-Poissonian state (SPS) with photon-number distribution
  $p_{\rm sP}$ [$p_{\rm a}$] arises in conditional photon-number resolving
  detection on the idler beam of TWB$_{\rm p}$ [TWB$_{\rm a}$]:
  \begin{eqnarray}  
    p_{\rm sP}(n_{\rm s};c_{\rm i}) &=& \sum_{n_{\rm i}=0}^{\infty}
    T(c_{\rm i},n_{\rm i};\eta,M_w,d)  p^{\rm TWB}_{\rm p}(n_{\rm s},n_{\rm i}),\nonumber \\
    p_{\rm a}(n'_{\rm s};\bar{c}_{\rm a}) &=& \sum_{\bar{n}_{\rm a}=0}^{\infty}
   T(\bar{c}_{\rm a},\bar{n}_{\rm a};\eta,\bar{M}_{\rm a},d)
    p^{\rm TWB}_{\rm a}(n'_{\rm s},\bar{n}_{\rm a}). \nonumber \\
    & & 
  \label{18}
  \end{eqnarray}
  The effective detection matrix $ T $ is given by Eq.~(3) of the main text. We note that the SPS with photon-number distribution $p_{\rm a}$ is used for PA.

  \item[PSTS] A state obtained by subtracting $ \bar{c}_{\rm s} $ photocounts from
   the TS given in Eq.~(\ref{17}) is obtained in the output of a beam splitter with transmissivity $t$ once $ \bar{c}_{\rm s} $ photocounts are detected in the
   other beam-splitter output monitored by a detector with the detection matrix $ T $:
   \begin{eqnarray}   
     p^{\rm s}_{\rm th}(n;\bar{c}_{\rm s}) &=& \sum_{\bar{n}_{\rm s}=0}^{\infty}
    \bar{T}_{\rm s}(\bar{c}_{\rm s},\bar{n}_{\rm s};\eta,\bar{M}_{\rm s},d)
     {\rm Bi}(n,n+ \bar{n}_{\rm s};t) 
    p_{\rm th}(n+\bar{n}_{\rm s});
    \label{19}
   \end{eqnarray}
  the binomial distribution Bi is given along the formula $ {\rm Bi}(n,m;t) = m! /
  [n! (m-n)!] \; t^{n}(1-t)^{m-n} $.

  \item[PATS] A state obtained by combining a sub-Poissonian field reached by post-selecting $\bar{c}_{\rm a}$
   photocounts with a TS is described by the convolution of photon-number distributions given in Eqs.~(\ref{17}) and (\ref{18}):
   \begin{eqnarray}  
    p^{\rm a}_{\rm th}(n;\bar{c}_{\rm a}) &=& \sum_{n'_{\rm s}=0}^{n} p_{\rm th}(n-n'_{\rm s}) p^{\rm a}(n'_{\rm s};\bar{c}_{\rm a}).
    \label{20}
   \end{eqnarray}

  \item[PSSPS] A state generated by subtracting $\bar{c}_{\rm s}$ photocounts, at a beam splitter
   with transmissivity $ t $ using a detector with detection matrix $ T $, from an SPS with the photon-number distribution given in Eq.~(\ref{18})
   is described by the formula:
   \begin{eqnarray}   
    p^{\rm s}_{\rm sP}(n;\bar{c}_{\rm s},c_{\rm i}) &=& \sum_{\bar{n}_{\rm s}=0}^{\infty}
    T(\bar{c}_{\rm s},\bar{n}_{\rm s};\eta,\bar{M}_{\rm s},d) {\rm Bi}(n,n+ \bar{n}_{\rm s};t) 
    p_{\rm sP}(n+\bar{n}_{\rm s};c_{\rm i}).
   \label{21}
   \end{eqnarray}

  \item[PASPS] A state obtained by combining a sub-Poissonian field reached by post-selecting $  \bar{c}_{\rm a}$ photocounts and the SPS with photon-number distribution written in  Eq.~(\ref{18}) is characterized by the following photon-number distribution:
   \begin{eqnarray}  
    p^{\rm a}_{\rm sP}(n;\bar{c}_{\rm a},c_{\rm i}) &=& \sum_{n'_{\rm s}=0}^{n} p_{\rm sP}(n-n'_{\rm s};c_{\rm i})
    p^{\rm a}(n'_{\rm s};\bar{c}_{\rm a}).\nonumber \\
   \label{22}
   \end{eqnarray}

  \item[PSTWB] Subtracting $\bar{c}_{\rm s}$ photocounts by the detector with the detection matrix $ T $ from the signal beam of twin beam TWB$ _{\rm p} $, we arrive
   at a PSTWB with the following joint photon-number distribution:
   \begin{eqnarray}   
    p^{\rm s}_{\rm TWB}(n_{\rm s},n_{\rm i};\bar{c}_{\rm s}) &=& \sum_{n'=0}^{\infty}
    T(\bar{c}_{\rm s},\bar{n}_{\rm s};\eta,\bar{M}_{\rm s},d) 
     {\rm Bi}(n_{\rm s},n_{\rm s}+ \bar{n}_{\rm s};t) p_{\rm p}^{\rm TWB}(n_{\rm s}+\bar{n}_{\rm s},n_{\rm i}).
    \label{23}
   \end{eqnarray}

  \item[PATWB] A state emerging by combining a sub-Poissonian field reached by post-selecting $ \bar{c}_{\rm a}$ photocounts  with the signal beam of twin beam TWB$ _{\rm p} $ has the following joint photon-number distribution:
   \begin{eqnarray}  
    p^{\rm a}_{\rm TWB}(n_{\rm s},n_{\rm i};\bar{c}_{\rm a}) &=& \sum_{n'_{\rm s}=0}^{n_{\rm s}}
    p_{\rm p}^{\rm TWB}(n_{\rm s}-n'_{\rm s},n_{\rm i})
    p^{\rm a}(n'_{\rm s};\bar{c}_{\rm a}).\nonumber \\
    \label{24}
   \end{eqnarray}
\end{description}

\bibliographystyle{iopart-num}
\bibliography{thapliyalSM}